\documentclass[a4paper,11pt]{article}
\pdfoutput=1 

\usepackage{jinstpub} 
\usepackage{tikz}
\usepackage{listings}
\usepackage{enumitem}

\newcommand{\e}[2]{$#1 \cdot 10^{#2}$}
\newcommand*{\doi}[1]{\href{https://doi.org/#1}{doi:\,#1}}
\newcommand*{\arxiv}[1]{\href{http://arxiv.org/abs/#1}{arXiv:\,#1}}

\title{\boldmath Achieving reliable UDP transmission at 10 Gb/s using BSD socket
for data acquisition systems}

\author[a,1]{M.J. Christensen,\note{Corresponding author.}}
\author[b]{T. Richter}

\affiliation[a]{European Spallation Source, Data Management and Software
   Centre, Ole Maal\o es vej 3, 2200 Copenhagen~N, Denmark}
\affiliation[b]{European Spallation Source ERIC, P.O. Box 176, 221 00 Lund, Sweden}

\emailAdd{mortenchristensen@esss.se}

\abstract{
User Datagram Protocol (UDP) is a commonly used protocol for data transmission in small embedded
systems. UDP as such is unreliable and packet losses can occur. The achievable data rates can suffer
if optimal packet sizes are not used. The alternative, Transmission Control Protocol (TCP) guarantees the
ordered delivery of data and automatically adjusts transmission to match the capability of the transmission
link. Nevertheless UDP is often favored over TCP due to its simplicity, small memory and instruction
footprints. Both UDP and TCP are implemented in all larger operating systems and commercial
embedded frameworks. In addition UDP also supported on a variety of small hardware platforms such as
Digital Signal Processors (DSP) Field Programmable Gate Arrays (FPGA). This is not so common for TCP.
This paper describes how high speed UDP based data transmission with very low packet error
ratios was achieved. The near-reliable communications link is used in a data acquisition (DAQ) system for
the next generation of extremely intense neutron source, European Spallation Source. This paper
presents measurements of UDP performance and reliability as achieved by employing several
optimizations. The measurements were performed on Xeon E5 based CentOS (Linux) servers. The
measured data rates are very close to the 10 Gb/s line rate, and zero packet loss was achieved.
The performance was obtained utilizing a single processor core as transmitter and a single core
as receiver. The results show that support for transmitting large data packets is a key parameter
for good performance.

Optimizations for throughput are: MTU, packet sizes, tuning Linux kernel parameters, thread affinity,
core locality and efficient timers.
}

\keywords{Computing (architecture, farms, GRID for recording, storage, archiving, and distribution
of data), Data acquisition concepts, Software architectures (event data models, frameworks and databases)}

\begin{document}
\maketitle
\flushbottom

\section{Introduction}
\label{sec:intro}
European Spallation Source \cite{ess1} is a next generation neutron source currently being developed
in Lund, Sweden. The facility will initially support about 16 different instruments for neutron
scattering. In addition to the instrument infrastructure, the ESS Data Management and Software
Centre (DMSC), located in Copenhagen, provides infrastructure and computational support for the acquisition,
event formation and long term storage of the experimental data.
  At the heart of each instrument is a neutron detector and its associated readout system. Both detectors
and readout systems are currently in the design phase and various prototypes have already been
produced \cite{ess2, ess3, ess4, ess5}.
  During experiments data is being produced at high rates: Detector data is read out by custom electronics
and the readings are converted into UDP packets by the readout system and sent to
event formation servers over 10 Gb/s optical Ethernet links.  The event formation servers
are based on general purpose CPUs and it is anticipated that most
if not all data reduction at ESS is done in software. This includes reception of raw readout data,
threshold rejection, clustering and event formation. For a detailed description of the software
architecture and early detector processing implementations see \cite{morten1}.
UDP is a simple protocol for connectionless
data transmission \cite{udp1} and packet loss can occur during transmission. Nevertheless
UDP is widely used, for example in the RD51 Scalable Readout System \cite{srs1}, or the
CMS trigger readout \cite{cms1}, both using 1 Gb/s Ethernet.
  The two central components are the readout system and the event formation
system. The readout system is a hybrid of analog and digital electronics. The electronics convert
deposited charges into electric signals which are digitized and timestamped. In the digital domain
simple data reduction such as zero suppression and threshold based rejection can be performed. The
event formation system receives these timestamped digital readouts and performs the necessary steps
to determine the position of the neutron. These processing steps are different for each detector
type.
   The performance of UDP over 10G Ethernet has been the subject of previous studies
\cite{bencivenni}  \cite{bortolotti}, which measured TCP and UDP performance
and CPU usages on Linux using commodity hardware. Both studies use a certain set of optimizations but
otherwise using standard Linux. In \cite{bencivenni} the transmitting process is found to be a
bottleneck in terms of CPU usage, whereas a comparison between Ethernet and InfiniBand \cite{bortolotti}
reinforces the earlier results and concludes that Ethernet is a serious contender for use in a readout
system. This study is aimed at characterizing the performance of a prototype data acquisition system
based on UDP. The study is not so much concerned with transmitter performance as we expect to receive
data from a FPGA based platform capable of transmitting at wire speed at all packet sizes. In stead
comparisons between the measured and theoretically possible throughput and measurements of packet
error ratios are presented. Finally, this paper presents strategies for optimizing the performance of
data transmission between the readout system and the event formation system.

\section{TCP and UDP pros and cons}
Since TCP is reliable and has good performance whereas UDP is unreliable why not always just use TCP?
The pros and cons for this will be discussed in the following. Both TCP and UDP are designed to provide end-to-end
communications between hosts connected over a network of packet forwarders. Originally these forwarders
were routers but today the group of forwarders include firewalls, load balancers, switches, Network Address
Translator (NAT) devices etc. TCP is connection oriented, whereas UDP is connectionless. This means
that TCP requires that a connection is setup before data can be transmitted. It also implies that
TCP data can only be sent from a single transmitter to a single receiver. In contrast UDP does not have a
connection concept and UDP data can be transmitted as either Internet Protocol (IP) broadcast or IP multicast.
  As mentioned earlier the main argument for UDP is that it is often supported on smaller systems
where TCP is not. A notable example are FPGA based systems (see \cite{fodisch1} for one case). For a
brief overview of efforts for providing TCP/IP support in FPGAs see \cite{wojciech1}.
But some of the TCP features are not actually improving the performance and reliability in the case of
special network topologies as explained below.

\subsection{Congestion}
Any forwarder is potentially subject to congestion and can drop packets
when unable to cope with the traffic load.
  TCP was designed to react to this congestion. Firstly
TCP has a slow start algorithm whereby the data rate is ramped up gradually in order not to
contribute to the network congestion itself. Secondly TCP will back off and reduce its transmission rate
when congestion is detected.
  In a readout system such as ours the network only consists of a data sender and a data receiver
with an optional switch connecting them. Thus the only places where congestion occurs are
at the sender or receiver. The readout system will typically produce data at near constant rates
during measurements so congestion at the receiver will result in reduced data rates by the
transmitter when using TCP. This first causes buffering at the transmitting application until the
buffer is full and eventually packets are lost.

For some detector readout it is not even evident that guaranteed delivery is necessary. In one detector
prototype we discarded around 24\% of the data due to threshold suppression, so spending extra time
making an occasional retransmission may not be worth the added complexity.

\subsection{Connections}
Since TCP requires the establishment of a connection, both the receiving and
transmitting applications must implement additional state to detect the possible loss of
a connection. For example upon reset of the readout system after a software upgrade or
a parameter change. With UDP the receiver will just 'listen' on a specified UDP port
whenever it is ready and receive data when it arrives. Correspondingly the transmitter
can send data whenever it is ready. UDP reception supports many-to-one communication,
supporting for example two or more readout systems in a single receiver. For TCP to support this would
require handling multiple TCP connections.

\subsection{Addressing}
UDP can be transmitted over IP as multicast. This means that a single sender can reach multiple
receivers without any additional programming effort. This can be used for seamless switchovers,
redundancy, load distribution, monitoring, etc.. Implementing this in TCP would add
complexity to the transmitter.

\vspace{5mm}
\noindent
In summary: For our purposes UDP appears to have more relevant features than TCP. Thus it is preferred
provided we can achieve the desired performance and reliability.

\section{Performance optimizations}
This section explains the factors that contribute to limiting
performance, reproducibility or accuracy of the measurements. Here we also discuss the
optimization strategies used to achieve the results.

\subsection{Transmission of data}
\label{datatrans}
\begin{figure}[htbp]
\centering
\includegraphics[width=.8\textwidth,clip]{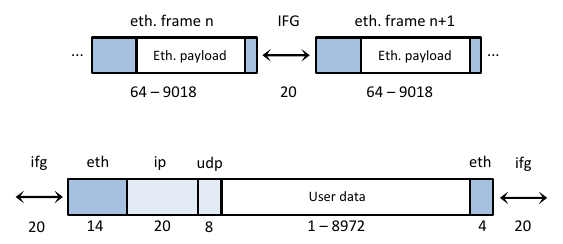}
\caption{\label{fig:headers} (top) Ethernet frames are separated by a 20 byte inter frame gap. (bottom)
The Ethernet, IP and UDP headers take up 46 bytes. The largest UDP user data size is 1472
bytes on most Ethernet interfaces due to a default MTU of 1500. This can be extended on
some equipment to 8972 bytes by the use of jumbo frames.}
\end{figure}

An Ethernet frame consists of a fixed 14 bytes header the Ethernet payload, padding and  a 4
byte checksum field. Padding is applied to ensure a minimum Ethernet packet size of 64 bytes.
There is a minimum gap between Ethernet frames of 20 bytes. This is called the Inter Frame Gap (IFG).
Standard Ethernet supports ethernet payloads from 1 to 1500 bytes. Ethernet frames with
payload sizes above 1500 bytes are called jumbo frames. Some Ethernet hardware support payload
sizes of 9000 bytes corresponding to Ethernet frame sizes of 9018 when including the header and checksum
fields. This is shown in Figure \ref{fig:headers} (top). The Ethernet payload consists of IP and UDP headers
as well as user data. This is illustrated in Figure \ref{fig:headers} (bottom).
  For any data to be transmitted over Ethernet, the factors influencing the packet and data rates
are: The link speed, IFG and the payload size. The largest supported Ethernet payload is called
the Maximum Transmission Unit (MTU). For further information see \cite{ethernet1} and \cite{ietf1}.

Sending data larger than the MTU will result in the data being split in
chunks of size MTU before transmission. Given a specific link speed and packet size,
the packet rate is given by
\begin{equation*}
  \text{rate} [\text{packets per second}] = \frac{\text{ls}}{8 \cdot (\text{ps} + \text{ifg})}
\end{equation*}
where {\em ls} is the link speed in b/s, {\em ps} the packet size and {\em ifg} the inter
frame gap. Thus for a 10 Gb/s Ethernet link, the packet rate for 64 byte packets is 14.88 M
packets per second (pps) as is shown in Table \ref{tab:i}.

\begin{table}[htbp]
  \centering
  \caption{\label{tab:i} Packet rates as function of packet sizes for 10 Gb/s Ethernet}
  \smallskip
  \begin{tabular}{|l|cccccccc|}
    \hline
      User data size [B] & 1  &       18 &   82 &  210 &  466 &   978  & 1472 & 8972 \\
    \hline
      Packet size [B]    & 64 &       64 &  128 &  256 &  512 &  1024  & 1518 & 9018 \\
    \hline
      Overhead [\%]      & 98.8 &   78.6 & 44.6 & 23.9 & 12.4 &   5.5  &  4.3 &  0.7 \\
    \hline
      Frame rate [Mpps]  & 14.88 & 14.88 & 8.45 & 4.53 & 2.35 &  1.20  & 0.81 & 0.14 \\
    \hline
  \end{tabular}
\end{table}

Packets arriving at a data acquisition system are subject to a nearly constant per-packet processing
overhead. This is due to interrupt handling, context switching, checksum validations and header
processing. At almost 15 M packets per second this processing alone can consume most of the available
CPU resources. In order to achieve maximum performance, data from the electronics readout should
be bundled into jumbo frames if at all possible. Using the maximum Ethernet packet size of 9018 bytes
reduces the per-packet overhead by a factor of 100. This does, however, come at the cost of larger
latency. For example the transmission time of 64 bytes + IFG is 67 ns, whereas for 9018 + IFG
it is 902 ns. For applications sensitive to latency a tradeoff must be made between low packet
rates and low latency.

Not all transmitted data are of interest for the receiver and can be considered as overhead. Packet
headers is such an example. The Ethernet, IP and UDP headers are always present and takes up a total
of 46 bytes as shown in Figure \ref{fig:headers} (bottom). The utilization of an Ethernet link can be calculated as
\begin{equation*}
  U = \frac{d}{d + 46 + \text{ifg} + \text{pad}}
\end{equation*} where {\em U} is the link utilization, {\em d} the user data size,
{\em ifg} the inter frame gap and {\em pad} is the padding mentioned earlier.
For user data larger than 18 bytes no padding is applied.
This means that for small user payloads the overhead can be significant, making it
impossible to achieve high throughput. For example transmitting a 32 bit counter over UDP will take
up 84 bytes on the wire (20 bytes IFG + 64 byte for a minimum Ethernet frame) and the overhead
will account for approx. 95\% of the available bandwidth. In contrast when sending 8972 byte user data
the overhead is as low as $0.73\%$.

\subsection{Network buffers and packet loss}
A UDP packet can be dropped in any part of the communications chain: The sender, the receiver, intermediate
systems such as routers, firewalls, switches, load balancers, etc. This makes it difficult in general to rely on UDP
for high speed communications. However for simple network topologies such as the ones found in
detector readout systems it is possible to achieve very reliable UDP communications. When for
example the system comprise two hosts (sender and receiver) connected via a switch of high quality,
the packet loss is mainly caused by the Ethernet Network Interface Card (NIC) transmit queue and the
socket receive buffer size.
Fortunately these can be optimized. The main parameters for controlling socket buffers are \texttt{rmem\_max}
and \texttt{wmem\_max}. The former is the size of the UDP socket receive buffer, whereas the latter is the
size of the UDP socket transmit buffer. To change these values from an application use \texttt{setsockopt()}, for example

\begin{verbatim}
int buffer = 4000000;
setsockopt(s, SOL_SOCKET, SO_SNDBUF, buffer, sizeof(buffer));
setsockopt(s, SOL_SOCKET, SO_RCVBUF, buffer, sizeof(buffer));
\end{verbatim}

In addition there is an internal queue for packet reception whose size (in packets) is named
\texttt{netdev\_max\_backlog}, and a network interface parameter, \texttt{txqueuelen} which were
also adjusted.

The default value of these parameters on Linux are not optimized for high speed data links such as
10 Gb/s Ethernet, so for this investigation the following parameters were used.

\begin{verbatim}
net.core.rmem_max=12582912
net.core.wmem_max=12582912
net.core.netdev_max_backlog=5000
txqueuelen 10000
\end{verbatim}

These values have largely been determined by experimentation although a detailed
description of these and other optimisations can be found in \cite{tcpperf1}. We also
configured the ethernet adaptors with an MTU of 9000 allowing user payloads up to 8972 bytes
when taking into account that IP and UDP headers are also transmitted.

\subsection{Core locality}
Modern CPUs rely heavily on cache memories to achieve performance. This holds
for both instructions and data access. For Xeon E5 processors there are three levels
of cache. Some is shared between instructions and data, some is dedicated. The L3
cache is shared across all cores and hyperthreads, whereas the L1 cache is only shared
between two hyperthreads. The way to ensure that the transmit and
receive applications always uses the same cache is to 'lock' the applications to
specific cores. For this we use the Linux command \texttt{taskset}.

This prevents the application processes to be moved to other cores and
thereby causing interrupts in the data processing, but it does not prevent other processes to be
swapped onto the same core.

\subsection{Timers}
The transmitter and receiver applications for this investigation periodically prints
out the measured data speed, PER and other parameters. Initially the standard  C++ chrono class timer
was used (version: libstdc++.so.6). But profiling showed that significant time was spent here, enough to affect the
measurements at high loads. Instead we decided to use the CPU's hardware based Time Stamp Counter
(TSC). TSC is a 64 bit counter running at CPU clock frequency. Since processor speeds are subject to
throttling, the TSC cannot be directly relied upon to
measure time. In this investigation time checking is a two-step process: First we estimate
when it is time to do the periodic update based on the inaccurate TSC value. Then we use the more expensive
C++ chrono functions to calculate the elapsed time used in the rate calculations. An example
of this is shown in the source code which is publicly available. See Section \ref{sec:code}
for instructions on how to obtain the source code.

\section{Testbed for the experiments}
\begin{figure}[htbp]
\centering
\includegraphics[width=.6\textwidth,clip]{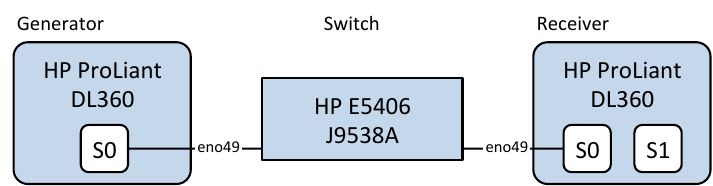}
\caption{\label{fig:expmt} Experimental setup.}
\end{figure}

The experimental configuration is shown in Figure \ref{fig:expmt}. It consists
of two hosts, one acting as a UDP data generator and the other as a UDP receiver.
The hosts are HPE ProLiant DL360 Gen9 servers connected to a 10 Gb/s Ethernet switch using
short (2 m) single mode fiber cables. The switch is a HP E5406 switch equipped with a J9538A 8-port SFP+ module. The
server specifications are shown in table \ref{tab:specs}. Except for processor internals the servers are
equipped with identical hardware.

\begin{table}[htbp]
  \small
  \centering
  \caption{\label{tab:specs} Hardware components for the testbed}
  \smallskip
  \begin{tabular}{|l|l|}
    \hline
      Motherboard                & HPE ProLiant DL360 Gen9 \\ \hline
      Processor type (receiver)  & Two 10-core Intel Xeon E5-2650v3 CPU @ 2.30GHz \\ \hline
      Processor type (generator) & One 6-core Intel Xeon E5-2620v3 CPU @ 2.40GHz  \\ \hline
      RAM                        & 64 GB (DDR4) - 4 x 16 GB DIMM - 2133MHz                      \\ \hline
      NIC                        & dual port Broadcom NetXtreme II BCM57810 10 Gigabit Ethernet \\ \hline
      Hard Disk                  & Internal SSD drive (120GB) for local installation of CentOS 7.1.1503 \\ \hline
      Linux kernel               & 3.10.0-229.7.2.el7.x86\_64 \\ \hline
  \end{tabular}
\end{table}

The data generator is a small C++ program using BSD socket, specifically the \texttt{sendto()}
system call for transmission of UDP data. The data receiver is based on a DAQ and
event formation system developed at ESS as a prototype. The system, named the Event Formation Unit (EFU),
supports loadable processing pipelines. A special UDP 'instrument' pipeline was created for the
purpose of these tests. Both the generator and receiver uses \texttt{setsockopt()} to adjust
transmit and receive buffer sizes.
Sequence numbers are embedded in the user payload by the transmitter allowing the receiver to detect
packet loss and hence to calculate packet error ratios.
Both the transmitting and receiving applications were locked to a specific processor core
using the \texttt{taskset} command and \texttt{pthread\_setaffinity\_np()} function.
The measured user payload data-rates were calculated using a combination of fast timestamp
counters and microsecond counters from the C++ chrono class.
Care was taken not to run other programs that might adversely
affect performance while performing the experiments. CPU usages were calculated from the
\texttt{/proc/stat} pseudofile as also used in \cite{bencivenni}.

A measurement series typically consisted of the following steps:
\begin{enumerate}
\item Start receiver
\item Start transmitter with specified packet size
\item Record packet error ratios (PER) and data rates
\item Stop transmitter and receiver after 400 GB
\end{enumerate}

The above steps were then repeated for measurements of CPU usage using \texttt{/proc/stat}
averaged over 10 second intervals.

A series of measurements of speed, packet error ratios and CPU usage where made as
a function of user data size for reasons discussed in Section \ref{datatrans}.

\subsection{Experimental limitations}
The current experiments are subject to some limitations. We do not however believe that
these pose any significant problems in the evaluation of the results. The main limitations
are described below.

\begin{description}[style=unboxed,leftmargin=0cm]
\item[Multi user issues:] The servers used for the tests are multi user systems in a
shared integration laboratory. Care was taken to ensure that other users were not running
applications at the same time to avoid competition for CPU, memory and network resources.
However a number of standard demon processes were running in the background, some of which
triggers the transmission of data and some of which are triggered by packet reception.

\item[Measuring affects performance:] Several configuration, performance and debugging tools need
access to kernel or driver data structures. Examples we encountered are \texttt{netstat},
\texttt{ethtool} and \texttt{dropwatch}. However the use of these tools can cause additional packet
drops when running at high system loads. These tools were not run while measuring packet losses.

\item[Packet reordering:] The test application is unable to detect misordered packets. Packet reordering
however is highly unlikely in the current setup, but would be falsely reported
as packet loss.

\item[Packet checksum errors:] The NICs perform checksums of Ethernet and IP in hardware. Thus
packets with wrong checksums will not be delivered to the application and subsequently
be falsely reported as packet loss. For the purpose of this study this is the
desired behavior.
\end{description}

\section{Performance}
The performance results covers user data speed, packet error ratios and CPU load. These topics will be covered
in the following sections.

\subsection{Data Speed}
The result of the measurements of achievable user data speeds is shown in Figure \ref{fig:combined} (a).
The figure shows both the measured and the theoretical maximum speed. For packets with user data
sizes larger than 2000 bytes the achieved rates match the theoretical maximum. However at smaller
data sizes the performance gap increases rapidly. It is clear that either the transmitter or the
receiver is unable to cope with the increasing load. This is mainly due to the higher
packet arrival rates occurring at smaller packet sizes. The higher rates increases the per-packet
overhead and also the number of interrupts and system calls.
  At the maximum data size of 8972 bytes the CPU load on the receiver was 20\%.

\subsection{Packet error ratios}
The achieved packet error ratios in this experiment are shown in Figure \ref{fig:combined} (b),
which also shows the corresponding values obtained using the default system parameters.
The raw measurements for the achieved values are listed in Table \ref{tab:ii}. It is observed
that the packet error ratios depends on the size of transmitted data. This
dependency is mainly caused by the per-packet overhead introduced by increasing packet rates with
decreasing size. The onset of packet loss coincides with the onset of deviation of observed
speed from the theoretical maximum speed suggesting a common cause.
No packet loss was observed for data larger than 2200 bytes. When packet loss
sets in at lower data sizes, the performance degrades rapidly: In the region from 2000 to 1700 bytes
the PER increases by more than four orders of magnitude from \e{1.3}{-6} to \e{7.1}{-2}.

\begin{table}[htbp]
  \footnotesize
  \centering
  \caption{\label{tab:ii} Packet error ratios as function of user data size}
  \smallskip
  \begin{tabular}{|l|c|c|c|c|c|c|c|c|}
    \hline
      size [B]&          64 &         128 &         256 &         472 &         772 &        1000  &        1472 &        1700 \\
    \hline
      PER     & \e{4.0}{-1} & \e{4.0}{-1} & \e{4.1}{-1} & \e{3.9}{-1} & \e{3.8}{-1} & \e{3.8}{-1}  & \e{2.0}{-1} & \e{7.1}{-2} \\
    \hline
    \hline
      size [B] &       1800 &        1900 &        2000 &        2200  & 2972 & 4472 & 5972 & 8972 \\
    \hline
      PER     & \e{3.2}{-3} & \e{6.1}{-6} & \e{1.3}{-6} &           0  &    0 &    0 &    0 &    0 \\
    \hline
  \end{tabular}
\end{table}

\subsection{CPU load}
\begin{figure}[htbp]
  \centering
  \includegraphics[width=1.0\textwidth,clip]{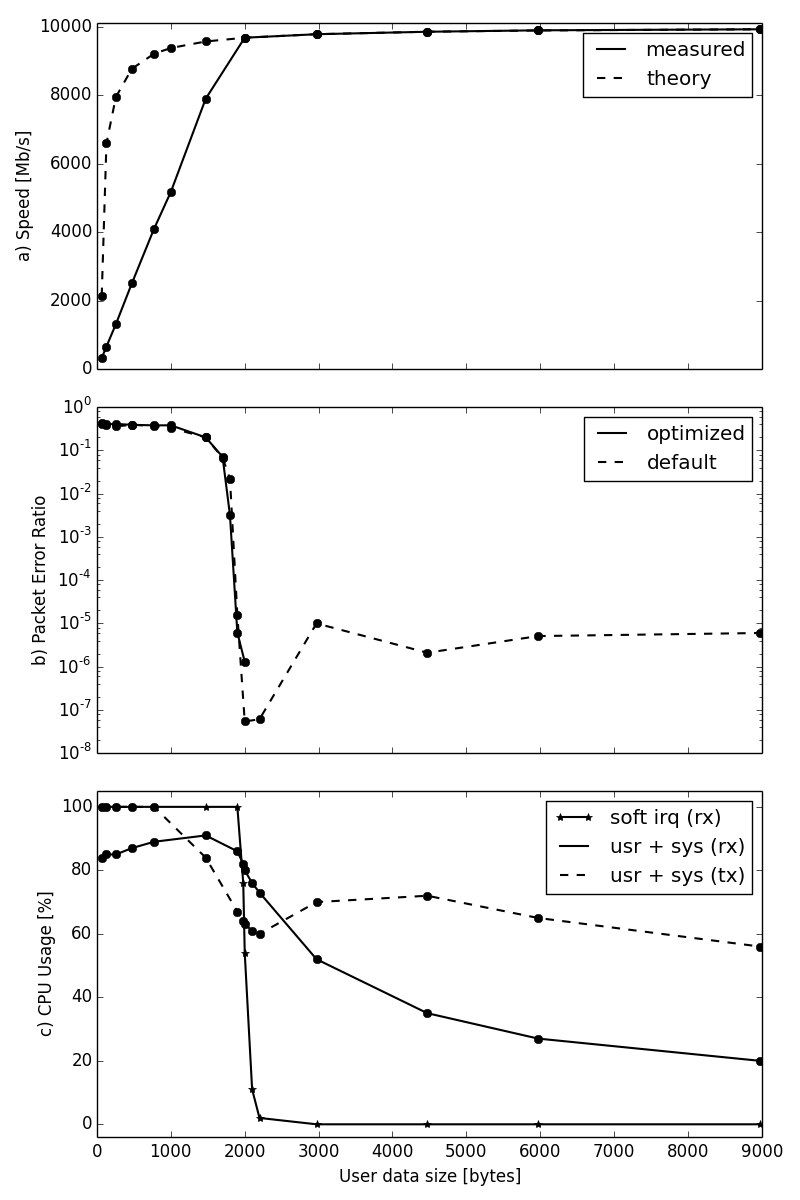}
  \caption{\label{fig:combined} Performance measurements. a) User data speed. b) Packet Error Ratio. c) CPU Load.
  Note that for the optimised values PER is zero for user data larger than or equal to 2200 bytes (solid line).}
\end{figure}

The CPU load as a function of user data size is shown in Figure \ref{fig:combined} (c). The observation
for both transmitter and receiver is that the CPU load increases with decreasing user data size.
When the transmitter reaches 100\% the receiver is slightly less busy at 84\%. There is a clear
cut-off value corresponding to packet loss and deviations from theoretical maximum speed around
user data sizes of 2000 bytes. The measured CPU loads indicate that the transmission is the
bottle neck at small data sizes (high packet rates), and that most CPU cycles are spent as System load
as also reported by \cite{bencivenni}.
  But the comparisons differ both qualitatively and quantitatively upon closer scrutiny. For example
in this study we find the total CPU load for the receiver (system + user) to be 20\% for user data sizes
of 8972 bytes. This is much lower than reported earlier. On the other hand we observe a sharp increase
CPU usage in soft IRQ from 0\% to 100\% over a narrow region which was not observed previously.
  We also observe a local minimum in Tx CPU load around 2000 bytes followed by a
rapid increase at lower data sizes.

\section{Conclusion}
Measurements of data rates and packet error ratios for UDP based communications at 10 Gb/s
have been presented. The data rates were achieved using standard hardware and software. No
modifications were made to the kernel network stack but some standard Linux commands were used
to optimise the behaviour of the system. The main change was increasing network buffers for
UDP communications from a small default value of 212 kB to 12 MB. In addition packet error
ratios were measured. The measurements shows that it is
possible to achieve zero packet error ratios at 10 Gb/s, but that this requires the use of large
Ethernet packets (jumbo frames), preferably as large as 9018 bytes. Thus the experiments have shown
that it is feasible to create a reliable UDP based data acquisition system supporting readout data
at 10 Gb/s.

This study supplements independent measurements done earlier \cite{bencivenni} and reveals
differences in performance across different platforms. The observed differences are likely to be caused
by differences in CPU generations, Ethernet NIC capabilities and Linux kernel versions. These differences
were not the focus of our study and have not been investigated further. But they do indicate that some
performance numbers are difficult to compare directly across different setups. They also provide a strong
hint to DAQ developers: When upgrading hardware or kernel versions in a (Linux based) DAQ system,
performance tests should be done to ensure that specifications are still met.

There are several ways to improve performance to achieve 10 Gb/s with smaller packet sizes,
but the complexity increases. For example it is possible to send and receive multiple messages using a
single system call such as \texttt{sendmmsg()} and \texttt{recvmmsg()} which will reduce
the number of system calls and should improve performance. It is also possible to use multiple
cores for the receiver instead of only one as we did in this test. This adds some complexity
that has to handle distributing packets across cores in case it cannot be done automatically. One method for
automatic load distribution is to use Receive Side Scaling (RSS). However this requires the
transmitter to use several different source ports in the UDP packet when transmitting instead of
one currently used. This may require changes to the readout system. It is also possible to move
network processing away from the kernel and into user space avoiding context switches, and to
change from interrupt driven reception to polling. These approaches are used in the
Intel Data Plane Development Kit (DPDK) software packet processing framework \cite{dpdk}
with impressive performance.

\appendix
\section{Source code}\label{sec:code}
The software for this project is released under a BSD license and is freely available at
GitHub \url{https://github.com/mortenjc/udpperf}. To build the programs used for these
experiments complete the following steps. To build and start the transmitter and receivers:
\begin{verbatim}
> git clone https://github.com/mortenjc/udpperf
> cd udpperf
> cmake -DCMAKE_BUILD_TYPE=Release ..
> make
> taskset -c coreid ./udptx -i ipaddress
> taskset -c coreid ./udprx
\end{verbatim}

  The programs have been demonstrated to build and run on maxOS (High Sierra, Catalina),
Ubuntu 16 and CentOS 7.1, however the \texttt{taskset} command is specific to Linux.

\section{System configuration}
The following commands were used (performed as superuser) to change the system parameters on CentOS.
Some of these have slightly different names on Ubuntu and generally are not available on macOS.
The examples below modifies network interface \emph{eno49}. This should be changed to match the name
of the interface on the actual system.
\begin{verbatim}
> sysctl -w net.core.rmem_max=12582912
> sysctl -w net.core.wmem_max=12582912
> sysctl -w net.core.netdev_max_backlog=5000
> ifconfig eno49 mtu 9000 txqueuelen 10000 up
\end{verbatim}

\acknowledgments
This work is funded by the EU Horizon 2020 framework, BrightnESS project 676548.

\vspace{5mm}
\noindent
We thank Sarah Ruepp, associate professor at DTU FOTONIK, and Irina Stefanescu, Detector Scientist
at ESS, for comments that greatly improved the manuscript.

\end{document}